\documentclass[12pt]{article}
\usepackage{times}
\usepackage{geometry,graphicx}
\usepackage[utf8]{inputenc}
\usepackage{enumitem,amssymb}
\usepackage{ragged2e}
\newlist{thematic}{itemize}{8}
\setlist[thematic]{label=$\square$}
\usepackage{pifont}
\usepackage[superscript,biblabel]{cite}

\def \chandra {{\em Chandra\ }}

\newcommand{\Msun}{\ifmmode M_{\odot} \else $M_{\odot}$\fi}
\def\arcsec{{\mbox{$^{\prime \prime}$}}}

\def\erg{{\rm\thinspace erg}}
\def\s{{\rm\thinspace s}}
\def\ergps{\hbox{$\erg\s^{-1}\,$}}

\def\micron{{\mbox{$\mu{\rm m}$}}}

\begin{document}
\raggedright
\huge
Astro2020 Science White Paper \linebreak

Black Hole Growth in Mergers and Dual AGN \linebreak
\normalsize

\noindent \textbf{Thematic Areas:} \hspace*{60pt} $\square$ Planetary Systems \hspace*{10pt} $\square$ Star and Planet Formation \hspace*{20pt}\linebreak
\makebox[0pt][l]{$\square$}\raisebox{.15ex}{\hspace{0.1em}$\checkmark$} Formation and Evolution of Compact Objects \hspace*{31pt} \makebox[0pt][l]{$\square$}\raisebox{.15ex}{\hspace{0.1em}$\checkmark$} Cosmology and Fundamental Physics \linebreak
  $\square$  Stars and Stellar Evolution \hspace*{1pt} $\square$ Resolved Stellar Populations and their Environments \hspace*{40pt} \linebreak
  \makebox[0pt][l]{$\square$}\raisebox{.15ex}{\hspace{0.1em}$\checkmark$}    Galaxy Evolution   \hspace*{45pt} \makebox[0pt][l]{$\square$}\raisebox{.15ex}{\hspace{0.1em}$\checkmark$}             Multi-Messenger Astronomy and Astrophysics \hspace*{65pt} \linebreak
  
\textbf{Principal Author:} 

Name:	Michael Koss
 \linebreak						
Institution: Eureka Scientific 
 \linebreak
Email: mike.koss@eurekasci.com 
 \linebreak
Phone:  206-372-1885
 \linebreak
 
\textbf{Co-authors:} Vivian U (University of California, Irvine), Edmund Hodges-Kluck (NASA GSFC), Ezequiel Treister (P. Universidad Catolica, Chile), Laura Blecha (University of Florida), Claudio Ricci (Universidad Diego Portales, Chile; Kavli Institute for Astronomy and Astrophysics, China), Jeyhan Kartaltepe (Rochester Institute of Technology), Dale Kocevski (Colby College), Julia M. Comerford (University of Colorado, Boulder), R. Scott Barrows (University of Colorado, Boulder), Claudia Cicone (INAF-OABrera), Francisco Muller-Sanchez (University of Memphis),  Kayhan G\"{u}ltekin (University of Michigan), Adi Foord (University of Michigan), Shobita Satyapal (Gorge Mason University), and Jennifer Lotz (Gemini Observatory)
  \linebreak

\justify

\textbf{Abstract  (optional):}
Hierarchical models of galaxy formation predict that galaxy mergers represent a significant transitional stage of rapid supermassive black hole (SMBH) growth. Yet, the connection between the merging process and enhanced active galactic nuclei (AGN) activity as well as the timescale of SMBH mergers remains highly uncertain. The breakthrough in reconciling the importance of galaxy mergers with black hole growth lies in a thoroughly-studied census of dual AGN across cosmic history, which will be enabled by next-generation observational capabilities, theoretical advances, and simulations. This white paper outlines the key questions in galaxy mergers, dual and offset AGN, and proposes multiwavelength solutions using future high-resolution observatories in the X-rays (\emph{AXIS, Lynx}), near and mid-infrared (30 meter class telescopes, \emph{JWST}), and submillimeter (ALMA).

\pagebreak

\section*{How Do Galaxy Mergers Lead to Black Hole Mergers?}

The general theory of large-scale structure formation predicts that galaxy mergers are a major component of galaxy growth. Since almost all massive galaxies at low redshift contain central supermassive black holes (SMBHs), it has long been predicted that when galaxies merge, so should their black holes.  However, the merging of SMBHs lacks direct observational evidence to-date. One such test, in addition to detection of gravitational waves (GWs), is to search for kpc-scale separation ``dual SMBHs’’ in nearby galaxies\cite{Steinborn:2016:1013}. Theoretical calculations indicate that a significant fraction of these sources are actively accreting\cite{VanWassenhove:2012:L7,Blecha:2013:2594}, and described as offset and dual AGN when one or both of the SMBH are accreting, respectively. \newline

Despite the intensive observational effort to search for dual and offset AGN\cite{Comerford:2009:956,Koss:2012:L22,Satyapal:2014:1297,MullerSanchez:2016:50}, we still do not know how common they are, particularly among a large sample of objects covering a wide range in BH mass, luminosity, and host galaxy properties. This is at least in part due to the enhanced obscuration of SMBHs in late-stage mergers\cite{Kocevski:2015:3629, Koss:2016:85,Ricci:2017:105}. Dual AGN are rare in the radio\cite{Burke-Spolaor:2011:2113} and optical selection techniques are rather inefficient, with a large fraction of false positives \cite{Liu:2010:L30,Fu:2011:103,Nevin:2016:67}. 
\newline

Based on existing observational samples of dual and offset AGN, there have been tantalizing hints that AGN triggering peaks in advanced-stage mergers where stellar bulge separations are $<$ 10 kpc \cite{Koss:2011:L42,Koss:2012:L22,Satyapal:2014:1297,Barrows:2017:27,Koss:2018:214a}, consistent with simulations of AGN accretion rates \cite{Stickley:2014:12}.  Additonally, dual AGN are preferentially triggered in major mergers, while offset AGN are triggered 
in minor mergers\cite{Koss:2012:L22,Comerford:2015:219,Barrows:2017:129,Shangguan:2016:50}. The offset AGN fraction does not depend on AGN luminosity, whereas the dual AGN fraction might\cite{Koss:2012:L22,Barrows:2017:129}, suggesting that major merger events are more efficient at instigating SMBH accretion than minor mergers.  However, a large and unbiased sample of dual and offset AGN is needed to confirm these trends and solidify an understanding of SMBH growth in mergers. \newline


With the recent discovery of GW emission from the merger of stellar mass black holes, interest in understanding the likely sources of GW emission from the merger of SMBHs has increased considerably.  In merger simulations, dynamical friction is thought to take the binary from galactic kpc scales (determined by the effective radius of the bulge, $R_\mathrm{eff}\approx 0.5$ kpc\cite{Dabringhausen:2008:864}) to $\approx$10 pc.  At 10 pc, stellar hardening may drive the binary to separations where it will be detectable by GW instruments ($10^{-2}$ pc to $10^{-4}$ pc separations).  Future space-based detectors such as {\it LISA}, would be sensitive to lower-mass SMBHs ($\sim10^5$--$10^7~\Msun$), making this mass range found in less massive galaxies a critical population to study.  Predictions of the expected detection rates for {\it LISA} are based on semi-analytic parameterizations of halo merger rates and the SMBH evolution \cite{Sesana:2018:42}, but are highly uncertain and vary by orders of magnitude.  In pulsar timing arrays, larger SMBHs ($>10^{8}\Msun$) at redshifts $z<$1 are expected to produce most of the signal \cite{Sesana:2009:L129,Mingarelli:2013}.  While the GW signal from these mergers will produce a stochastic background of confused sources, the rarest brightest nearby sources may dominate the signal \cite{Ravi:2014:56,Roebber:2016:163}.  The study of kpc to sub-kpc mergers is therefore critical for comparison with cosmological merger-rate models, because it can help constrain the timescales for SMBH inspiral and the rate of such events.   In particular, cosmological hydrodynamics simulations are being used to make predictions that, unlike semi-analytic models, can account for the internal dynamics of galaxies and their SMBHs \cite{Habouzit:2016:1901,Rosas-Guevara:2016:190,Kelley:2017:3131,Kelley:2017:4508,Kelley:2018:964}. Next-generation simulations will probe smaller-scale dynamics with higher resolution and more accurate sub-grid models that can only be tested with higher resolution models.\newline  

Another important aspect of the merger model is the prevalance of ``recoiling'' black holes. SMBHs are generally thought to sit at the center of their host galaxies.  However, recent simulations have predicted that “wandering” SMBHs might exist within their host galaxies for several Gyr \cite{Tremmel:2017:1121}.  “Recoiling” black holes have also been predicted because of GWs being emitted more strongly in one direction \cite{Detweiler:1983:67,Gualandris:2005:845}.  The strength of the kick causing the recoil will depend on the rate and direction of spin of both black holes--providing important insights into the types of black hole mergers that occur in the universe.  Thus, the ability to constrain kicked AGN using high spatial resolution observations of both the host galaxy and AGN is critical.

\section*{How Do Dual AGN Affect Galaxies?}

Since dual AGNs seem to be bright at small distances, most of the merger-triggered SMBH growth may occur in these systems. There is now ample evidence that such merger-triggered growth drives outflows from galaxies and otherwise impacts their ISM \cite{Cicone:2018:143}, but there is little observational evidence for how the presence (and separation) of a dual AGN affects this process, as the ISM surrounding each SMBH may be affected by the other, and the angular momentum of the gas must be accounted for in the merger process. The next step beyond identifying dual AGNs is to determine how, as a population, they impact their surroundings on scales of 100~pc to 1~kpc.

\section*{Open Questions}

Over the next decade, high spatial resolution observations will enable the most significant progress in understanding galaxy mergers, dual AGN, and offset AGN along with their frequency and environment throughout cosmic time.  Some key questions include:
\begin{enumerate}
\item What are timescales for sub-kpc binary inspiral and how does the galaxy merger rate translate to a SMBH merger rate?
\item What are the frequency, environment, and luminosity dependence of dual AGN, and offset AGN, and how does it evolve with redshift?
\item What is the relation between SMBH growth and the gas (inflow and outflow) and stellar dynamics in galaxy mergers? 

\end{enumerate}

\section*{Observational Requirements and Prospects}
In order to fully investigate black hole growth in galaxy mergers, high resolution observations provide the greatest opportunity to study SMBH growth in dual and offset AGN, as well as the impact of this growth on the galaxy (i.e., how important is feedback during the dual AGN phase?). Here we sketch how current and future observational capabilities will follow dual AGN from their formation at $\sim$10~kpc down to where the constituent SMBHs form a binary at a few pc.

\begin{enumerate}[topsep=2mm,fullwidth,itemindent=2mm]

\item {\bf Wide-Field X-ray Surveys (1,000-10,000 pc)}: X-rays, especially hard X-rays at E$>$10 keV (rest-frame), are very efficient at detecting AGNs because the sky is otherwise very dark. All but one of the \textit{bona fide} dual AGN currently known\cite{Rodriguez:2006:49} are separated by more than 3~kpc and have luminosities $L_X>10^{42.5}$ \ergps\ where they could be detected in the X-rays to $z\gtrsim2$, provided sufficient resolution (at $z=1$, 1\arcsec\ corresponds to 8~kpc). High-resolution X-ray images are thus one of the best ways to identify dual AGN, but \chandra\ has too little sensitivity and a small field of view where the resolution is better than 1\arcsec, so the number of dual AGN expected even in very deep fields is small. This is unfortunate, as a large sample is required to understand how/when they are triggered, their lifetimes, and how they connect to SMBH growth. This motivates a wide-field ($>$15~arcmin field of view), high-resolution ($<$1\arcsec) X-ray camera that could detect thousands of dual AGN (e.g., \textit{Lynx} or \textit{AXIS}). Importantly, a search for dual AGN is contiguous with wide, medium, and deep surveys to search for high-$z$ AGN (Fig.~1), as such surveys will capture many luminous AGN out to $z=2$. Over a 5-year mission, this would allow a blind search for dual and offset AGN in $\sim$750,000 serendipitous AGN, and enable measurement of the dual or offset AGN frequency, environment, luminosity dependence, and correlation with obscuring column. Many of these candidates will be formally unresolved, as the dual fraction and the luminosity rises with decreasing separation, but recent techniques\cite{Foord:2019} have been found to efficiently characterize the dual AGN fraction to $\lesssim$0.4\arcsec with \textit{Chandra}. 

\begin{figure}
\centering
\raisebox{0.65cm}{\includegraphics[width=6cm]{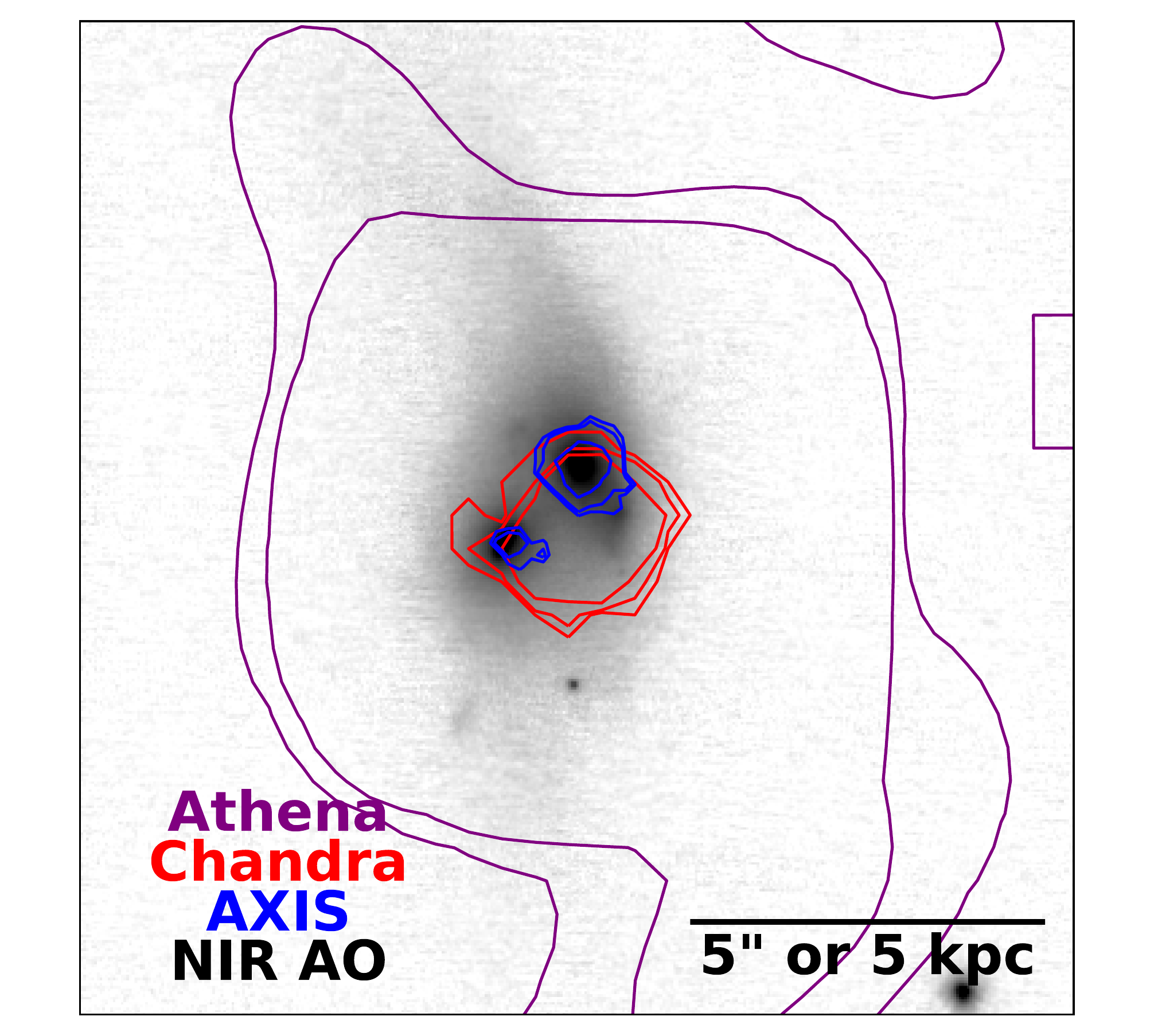}}
\includegraphics[width=10cm]{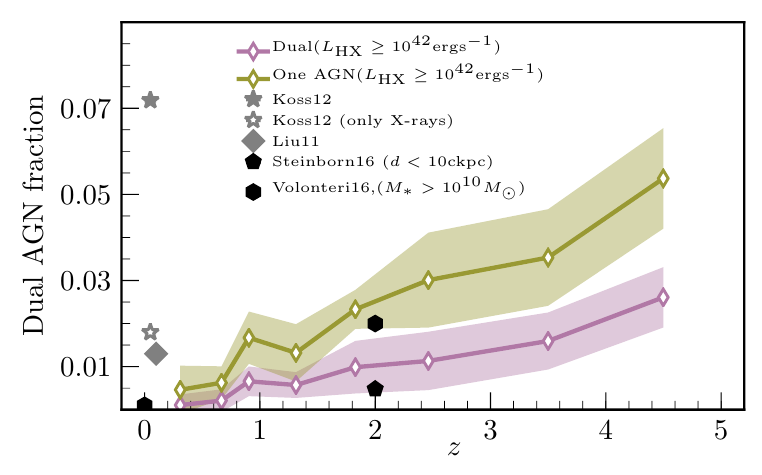}\
\vspace{-0.25cm}
\caption{\small{{\em Left}: \textbf{High-resolution NIR and X-ray imaging see through enshrouding dust to detect dual AGN with $<$1~kpc separation} out to high redshifts. The combination of the two is ideal for verifying dual AGN candidates. Simulations show that \textit{AXIS} (blue contours) resolves the dual AGNs in this late-stage merger, whereas \textit{Athena} (purple contours) does not; \chandra\ (red contours) simply lacks the sensitivity. {\em Right}: \textbf{A wide-field, high-resolution X-ray survey will measure the dual AGN fraction across cosmic time} and provide a direct constraint on models for the merger history of BHs. Cosmological simulations\cite{Rosas-Guevara:2016:190} predict that the fraction of dual AGN increases dramatically with redshift, but a \chandra study of nearby dual AGN\cite{Koss:2012:4264} finds a much higher than expected fraction. }}

\end{figure}

The large parameter space explored by these surveys (which is difficult to match in deep pencil-beam surveys) will produce the clearest view of dual and offset AGN with kpc~scale separations, answering the questions of how common and luminous they are, and how this depends on redshift. These data will constrain the SMBH merger function for the expected $LISA$ event rates and pulsar timing array signals. 

\item {\bf Optical and infrared identification (100-1,000~pc)}: 
Recent near-infrared (NIR) observations revealed signatures of late-stage mergers in a large fraction of the host galaxies of obscured, luminous AGN (Fig.~2)\cite{Koss:2018:214a}. This suggests that closely separated, dual AGN may be common, but detecting them requires high resolution telescopes that can peer through enshrouding dust. 

\begin{figure}
\centering
\includegraphics[width=10cm]{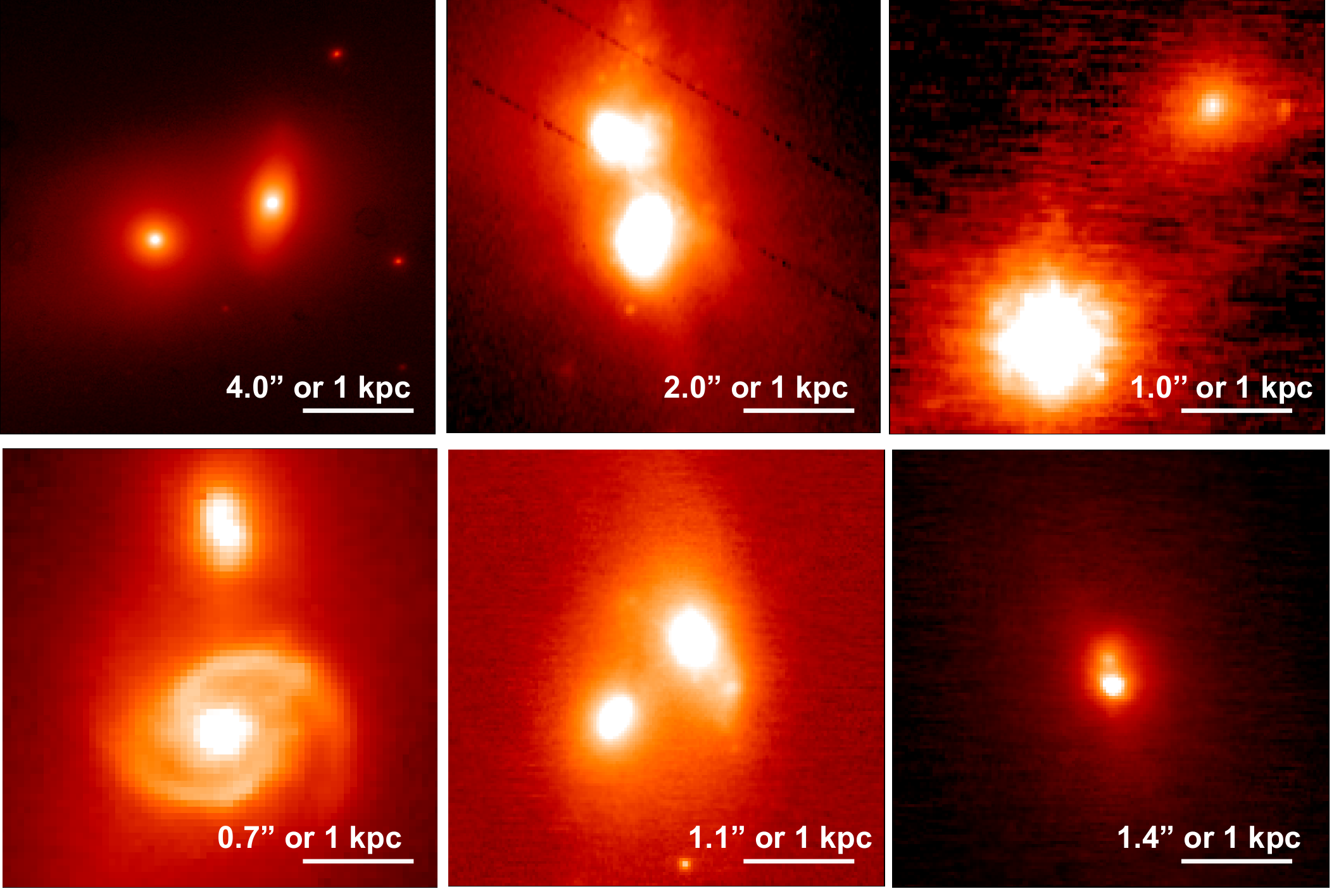}
\caption{\small{ \textbf{A NIR survey of luminous AGNs demonstrates the power to resolve closely separated nuclei.} These $Kp$-band (2.1 \micron) images of nuclear mergers from the Keck/NIRC2 instrument\cite{Koss:2018:214a} are 4x4~kpc in size. Future 30m telescopes and JWST will find many more such systems out to high redshifts ($z\sim1$).}}
\label{xraydualfrac}
\end{figure}

Extremely Large Telescopes (ELTs) will facilitate the discovery of dual AGN in the optical and IR bands at cosmological distances. The electromagnetic signatures of dual SMBHs rely both on spatially resolving the candidates and confirming the AGN nature of each, using kinematics and/or well-established line ratio diagnostics\cite{U13,U19}. With plate scales of milli-arcseconds per pixel, ELTs can probe projected separations of $\sim$100~pc to $z$=2, thereby completing the census of dual AGN up to cosmic noon. The same plate scale can resolve $\sim$5~pc at $d$=100~Mpc, detecting mergers down to near the ``final parsec'' in nearby galaxies.  The \textit{James Webb Space Telescope} (JWST) will also soon provide sub-kpc imaging and spectroscopy in the IR that will find dual and offset AGN candidates and close mergers at redshifts beyond the volume currently accessible to \textit{Hubble}:   NIRCam can detect late-stage mergers in AGN hosts out to $z\sim 2$, while MIRI (at 5-21~$\mu$m) can detect obscured AGN using the colors $S_{15}/S_{7.7}$ vs. $S_{18}/S_{10}$ at $z= 1$-2 via reprocessed dust emission.  While existing IR selection techniques using \textit{Spitzer} and WISE \cite{Donley:2012:142} require hot dust emission to overpower the underlying stellar emission from the host galaxy for an AGN to be identified, simulations show that MIRI colors can identify AGN accreting at low Eddington ratios ($\lambda_{\rm Edd}$) in normal star-forming galaxies (where the nuclear emission does not dominate the bolometric luminosity)\cite{Kirkpatrick:2017:111}.  

Candidates selected by imaging can be followed up by spectroscopy. MIRI/MRS $+$ NIRSpec IFU will be a powerful tool to identify buried AGNs, characterize the state of the molecular and ionized gas, search for AGN and starburst-driven outflows, and obtain kinematic information from the mergers, connecting the presence of dual AGN to their environments. Meanwhile, medium-resolution integral-field spectrographs on ELTs will reveal the chemical compositions and kinematics of the stars and nebula within 100~pc of the nucleus, incorporating well-established optical diagnostics for AGN ionization.

\item {\bf Detailed Sub-mm Observations ($<$100~pc)}: 
Sensitive (sub-)mm arrays such as ALMA have enabled high-resolution studies of the kinematic and physical properties of the cold interstellar medium in late-stage galaxy mergers, including dual AGNs and dual AGN candidates (e.g. NGC6240\cite{Scoville:2015:70,Cicone:2018:143,Saito:2018:L52}, Mrk463\cite{Treister:2018:83}, Arp220\cite{Scoville:2015:70,Scoville:2017:66,Sakamoto:2017:14,Barcos-Munoz:2018:L28}; NGC3256\cite{Sakamoto:2014:90}). Similar observations of a wider sample are necessary to detect most of the gas mass (in molecular clouds) and thus understand how SMBHs grow in mergers, and how this growth affects the galaxy.

\begin{figure}[t]
\begin{center}
\includegraphics[scale=0.3]{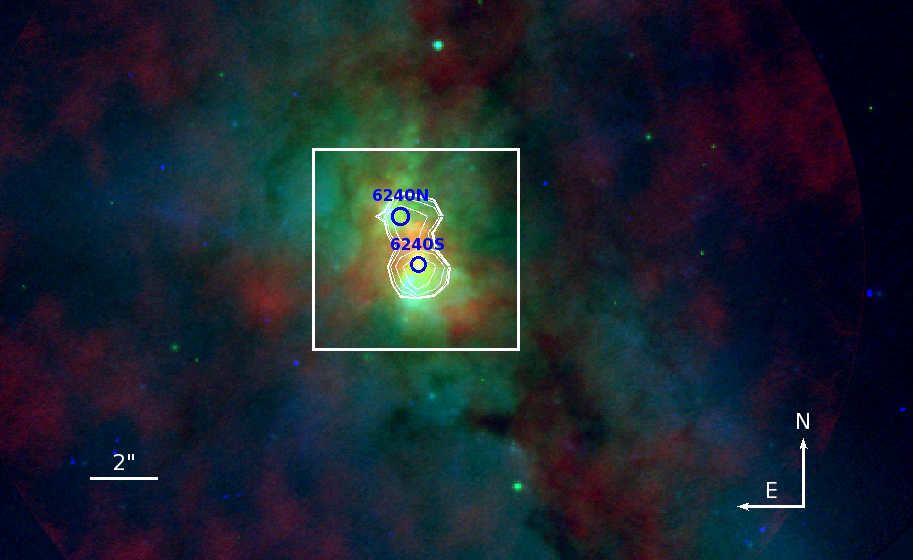}
\includegraphics[scale=0.17]{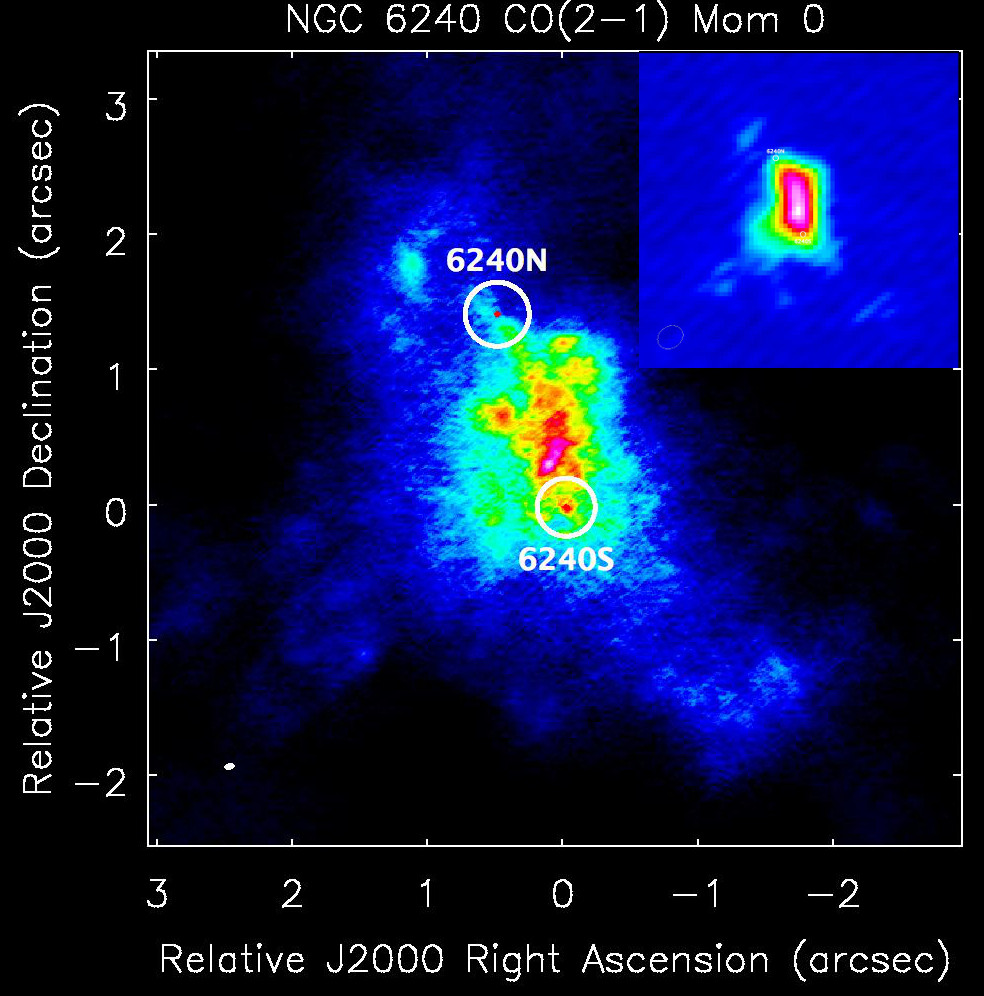}
\end{center}
\vspace{-0.5cm}
\caption{\small{{\em Left}: \textbf{Combined ALMA and HST observations of NGC6240 reveal the gas dynamics around the rapidly growing, dual SMBH}. The $^{12}$CO(2-1) intensity is shown in red, while 
while green and blue correspond to the F814W and F450W HST bands, respectively. White contours show the 230 GHz continuum emission. {\em Right}: \textbf{A zoomed-in view with ALMA resolves the sphere of influence of each SMBH, enabling a direct study of how galaxy growth and BH growth are linked.} The white circles centered on each nuclei show the sphere of influence of each SMBH, relative to the gas at very high resolution. For comparison, the inset shows past observations of the same region at 0.4$''$ resolution \cite{Treister:2019}.}}
\label{ngc6240_alma_mom0}
\end{figure} 

In the local Universe, major mergers tend to coincide with ultra-luminous IR galaxies (ULIRGs), which have extreme molecular ISM excitations and kinematic conditions that clearly set them apart from the (non-merging) star forming galaxy population\cite{Aalto:1995:369,Downes:1998:615,Papadpoulos:2012:2601}. However, previous interferometric studies were affected on the one hand by a limited angular resolution and on the other hand by limited {\it (u,v)} coverage, leading to poor sensitivity to large-scale structures. ALMA has significantly improved on these limitations and reaches angular resolutions of $\sim$0.03$''$ with the most extended baselines, as can be seen from the ALMA observations of NGC6240 shown in Fig.\ref{ngc6240_alma_mom0}. This makes it possible to probe the molecular medium within the inner 10s of pc in the nuclei of ULIRGs and dual AGN. For example, in the case of the dual AGN NGC6240, ALMA observations of CO, [CI], and other lines pinpointed the regions with the most extreme ISM excitation and linked them to a massive molecular outflow\cite{Cicone:2018:143}. 
ALMA can extend this pioneering study to other closely separated dual AGNs (identified through the techniques above), thereby revealing how the late-stage merger process relates to SMBH growth and galaxy-scale feedback. \newline
\end{enumerate}
\vspace{-0.5cm}

\noindent The combination of future high-resolution observatories in the X-rays (\emph{AXIS, Lynx}), near and mid-infrared (30 meter class telescopes, \emph{JWST}), and submillimeter (ALMA) is critical to trace black hole growth in galaxy mergers and dual AGN from their formation at kpc scales to their culmination in black-hole binaries at pc scales. 
\pagebreak
\noindent 
\bibliography{bibfinal.bib}

\begin{thebibliography}{10}
\expandafter\ifx\csname url\endcsname\relax
  \def\url#1{\texttt{#1}}\fi
\expandafter\ifx\csname urlprefix\endcsname\relax\def\urlprefix{URL }\fi
\providecommand{\bibinfo}[2]{#2}
\providecommand{\eprint}[2][]{\url{#2}}

\bibitem{Steinborn:2016:1013}
\bibinfo{author}{Steinborn, L.~K.} \emph{et~al.}
\newblock \bibinfo{title}{{Origin and properties of dual and offset active
  galactic nuclei in a cosmological simulation at z=2}}.
\newblock \emph{\bibinfo{journal}{Mon. Not. R. Astron. Soc.}}
  \textbf{\bibinfo{volume}{458}}, \bibinfo{pages}{1013--1028}
  (\bibinfo{year}{2016}).

\bibitem{VanWassenhove:2012:L7}
\bibinfo{author}{Van~Wassenhove, S.} \emph{et~al.}
\newblock \bibinfo{title}{{Observability of Dual Active Galactic Nuclei in
  Merging Galaxies}}.
\newblock \emph{\bibinfo{journal}{Astrophys. J. Lett.}}
  \textbf{\bibinfo{volume}{748}}, \bibinfo{pages}{L7} (\bibinfo{year}{2012}).

\bibitem{Blecha:2013:2594}
\bibinfo{author}{Blecha, L.}, \bibinfo{author}{Loeb, A.} \&
  \bibinfo{author}{Narayan, R.}
\newblock \bibinfo{title}{{Double-peaked narrow-line signatures of dual
  supermassive black holes in galaxy merger simulations}}.
\newblock \emph{\bibinfo{journal}{Mon. Not. R. Astron. Soc.}}
  \textbf{\bibinfo{volume}{429}}, \bibinfo{pages}{2594--2616}
  (\bibinfo{year}{2013}).

\bibitem{Comerford:2009:956}
\bibinfo{author}{Comerford, J.~M.} \emph{et~al.}
\newblock \bibinfo{title}{{Inspiralling Supermassive Black Holes: A New
  Signpost for Galaxy Mergers}}.
\newblock \emph{\bibinfo{journal}{Astrophys. J.}}
  \textbf{\bibinfo{volume}{698}}, \bibinfo{pages}{956--965}
  (\bibinfo{year}{2009}).

\bibitem{Koss:2012:L22}
\bibinfo{author}{Koss, M.} \emph{et~al.}
\newblock \bibinfo{title}{{UNDERSTANDING DUAL ACTIVE GALACTIC NUCLEUS
  ACTIVATION IN THE NEARBY UNIVERSE}}.
\newblock \emph{\bibinfo{journal}{Astrophys. J.}}
  \textbf{\bibinfo{volume}{746}}, \bibinfo{pages}{L22} (\bibinfo{year}{2012}).

\bibitem{Satyapal:2014:1297}
\bibinfo{author}{Satyapal, S.} \emph{et~al.}
\newblock \bibinfo{title}{{Galaxy pairs in the Sloan Digital Sky Survey - IX.
  Merger-induced AGN activity as traced by the Wide-field Infrared Survey
  Explorer}}.
\newblock \emph{\bibinfo{journal}{Mon. Not. R. Astron. Soc.}}
  \textbf{\bibinfo{volume}{441}}, \bibinfo{pages}{1297--1304}
  (\bibinfo{year}{2014}).

\bibitem{MullerSanchez:2016:50}
\bibinfo{author}{M{\"u}ller~S{\'a}nchez, F.}, \bibinfo{author}{Comerford, J.},
  \bibinfo{author}{Stern, D.} \& \bibinfo{author}{Harrison, F.~A.}
\newblock \bibinfo{title}{{The Nature of Active Galactic Nuclei with Velocity
  Offset Emission Lines}}.
\newblock \emph{\bibinfo{journal}{Astrophys. J.}}
  \textbf{\bibinfo{volume}{830}}, \bibinfo{pages}{50} (\bibinfo{year}{2016}).

\bibitem{Kocevski:2015:3629}
\bibinfo{author}{Kocevski, D.~D.} \emph{et~al.}
\newblock \bibinfo{title}{{Are Compton-Thick AGN the Missing Link Between
  Mergers and Black Hole Growth?}}
\newblock \emph{\bibinfo{journal}{arXiv.org}} \bibinfo{pages}{3629}
  (\bibinfo{year}{2015}).
\newblock \eprint{1509.03629}.

\bibitem{Koss:2016:85}
\bibinfo{author}{Koss, M.~J.} \emph{et~al.}
\newblock \bibinfo{title}{{A New Population of Compton-thick AGNs Identified
  Using the Spectral Curvature above 10 keV}}.
\newblock \emph{\bibinfo{journal}{Astrophys. J.}}
  \textbf{\bibinfo{volume}{825}}, \bibinfo{pages}{85} (\bibinfo{year}{2016}).

\bibitem{Ricci:2017:105}
\bibinfo{author}{Ricci, C.} \emph{et~al.}
\newblock \bibinfo{title}{{NuSTAR Observations of WISE J1036+0449, a Galaxy at
  z{\textasciitilde}1 Obscured by Hot Dust}}.
\newblock \emph{\bibinfo{journal}{Astrophys. J.}}
  \textbf{\bibinfo{volume}{835}}, \bibinfo{pages}{105} (\bibinfo{year}{2017}).

\bibitem{Burke-Spolaor:2011:2113}
\bibinfo{author}{Burke-Spolaor, S.}
\newblock \bibinfo{title}{{A radio Census of binary supermassive black holes}}.
\newblock \emph{\bibinfo{journal}{Mon. Not. R. Astron. Soc.}}
  \textbf{\bibinfo{volume}{410}}, \bibinfo{pages}{2113--2122}
  (\bibinfo{year}{2011}).

\bibitem{Liu:2010:L30}
\bibinfo{author}{Liu, X.}, \bibinfo{author}{Greene, J.~E.},
  \bibinfo{author}{Shen, Y.} \& \bibinfo{author}{Strauss, M.~A.}
\newblock \bibinfo{title}{{Discovery of Four kpc-scale Binary Active Galactic
  Nuclei}}.
\newblock \emph{\bibinfo{journal}{Astrophys. J. Lett.}}
  \textbf{\bibinfo{volume}{715}}, \bibinfo{pages}{L30--L34}
  (\bibinfo{year}{2010}).

\bibitem{Fu:2011:103}
\bibinfo{author}{Fu, H.}, \bibinfo{author}{Myers, A.~D.},
  \bibinfo{author}{Djorgovski, S.~G.} \& \bibinfo{author}{Yan, L.}
\newblock \bibinfo{title}{{Mergers in Double-peaked [O III] Active Galactic
  Nuclei}}.
\newblock \emph{\bibinfo{journal}{Astrophys. J.}}
  \textbf{\bibinfo{volume}{733}}, \bibinfo{pages}{103} (\bibinfo{year}{2011}).

\bibitem{Nevin:2016:67}
\bibinfo{author}{Nevin, R.}, \bibinfo{author}{Comerford, J.},
  \bibinfo{author}{M{\"u}ller~S{\'a}nchez, F.}, \bibinfo{author}{Barrows, R.}
  \& \bibinfo{author}{Cooper, M.}
\newblock \bibinfo{title}{{The Origin of Double-peaked Narrow Lines in Active
  Galactic Nuclei. II. Kinematic Classifications for the Population at $z <
  0.1$}}.
\newblock \emph{\bibinfo{journal}{Astrophys. J.}}
  \textbf{\bibinfo{volume}{832}}, \bibinfo{pages}{67} (\bibinfo{year}{2016}).

\bibitem{Koss:2011:L42}
\bibinfo{author}{Koss, M.} \emph{et~al.}
\newblock \bibinfo{title}{{Chandra Discovery of a Binary Active Galactic
  Nucleus in Mrk 739}}.
\newblock \emph{\bibinfo{journal}{Astrophys. J. Lett.}}
  \textbf{\bibinfo{volume}{735}}, \bibinfo{pages}{L42} (\bibinfo{year}{2011}).

\bibitem{Barrows:2017:27}
\bibinfo{author}{Barrows, R.~S.}, \bibinfo{author}{Comerford, J.~M.},
  \bibinfo{author}{Zakamska, N.~L.} \& \bibinfo{author}{Cooper, M.~C.}
\newblock \bibinfo{title}{{Observational Constraints on Correlated Star
  Formation and Active Galactic Nuclei in Late-stage Galaxy Mergers}}.
\newblock \emph{\bibinfo{journal}{Astrophys. J.}}
  \textbf{\bibinfo{volume}{850}}, \bibinfo{pages}{27} (\bibinfo{year}{2017}).

\bibitem{Koss:2018:214a}
\bibinfo{author}{Koss, M.~J.} \emph{et~al.}
\newblock \bibinfo{title}{{A population of luminous accreting black holes with
  hidden mergers}}.
\newblock \emph{\bibinfo{journal}{Nature}} \textbf{\bibinfo{volume}{563}},
  \bibinfo{pages}{214--216} (\bibinfo{year}{2018}).

\bibitem{Stickley:2014:12}
\bibinfo{author}{Stickley, N.~R.} \& \bibinfo{author}{Canalizo, G.}
\newblock \bibinfo{title}{{Stellar Velocity Dispersion in Dissipative Galaxy
  Mergers with Star Formation}}.
\newblock \emph{\bibinfo{journal}{Astrophys. J.}}
  \textbf{\bibinfo{volume}{786}}, \bibinfo{pages}{12} (\bibinfo{year}{2014}).

\bibitem{Comerford:2015:219}
\bibinfo{author}{Comerford, J.~M.} \emph{et~al.}
\newblock \bibinfo{title}{{Merger-driven Fueling of Active Galactic Nuclei: Six
  Dual and Offset AGNs Discovered with Chandra and Hubble Space Telescope
  Observations}}.
\newblock \emph{\bibinfo{journal}{Astrophys. J.}}
  \textbf{\bibinfo{volume}{806}}, \bibinfo{pages}{219} (\bibinfo{year}{2015}).

\bibitem{Barrows:2017:129}
\bibinfo{author}{Barrows, R.~S.}, \bibinfo{author}{Comerford, J.~M.},
  \bibinfo{author}{Greene, J.~E.} \& \bibinfo{author}{Pooley, D.}
\newblock \bibinfo{title}{{Spatially Offset Active Galactic Nuclei. II.
  Triggering in Galaxy Mergers}}.
\newblock \emph{\bibinfo{journal}{Astrophys. J.}}
  \textbf{\bibinfo{volume}{838}}, \bibinfo{pages}{129} (\bibinfo{year}{2017}).

\bibitem{Shangguan:2016:50}
\bibinfo{author}{Shangguan, J.} \emph{et~al.}
\newblock \bibinfo{title}{{Chandra X-Ray and Hubble Space Telescope Imaging of
  Optically Selected Kiloparsec-scale Binary Active Galactic Nuclei. II. Host
  Galaxy Morphology and AGN Activity}}.
\newblock \emph{\bibinfo{journal}{Astrophys. J.}}
  \textbf{\bibinfo{volume}{823}}, \bibinfo{pages}{50} (\bibinfo{year}{2016}).

\bibitem{Dabringhausen:2008:864}
\bibinfo{author}{{Dabringhausen}, J.}, \bibinfo{author}{{Hilker}, M.} \&
  \bibinfo{author}{{Kroupa}, P.}
\newblock \bibinfo{title}{{From star clusters to dwarf galaxies: the properties
  of dynamically hot stellar systems}}.
\newblock \emph{\bibinfo{journal}{\mnras}} \textbf{\bibinfo{volume}{386}},
  \bibinfo{pages}{864--886} (\bibinfo{year}{2008}).
\newblock \eprint{0802.0703}.

\bibitem{Sesana:2018:42}
\bibinfo{author}{Sesana, A.}, \bibinfo{author}{Haiman, Z.},
  \bibinfo{author}{Kocsis, B.} \& \bibinfo{author}{Kelley, L.~Z.}
\newblock \bibinfo{title}{{Testing the Binary Hypothesis: Pulsar Timing
  Constraints on Supermassive Black Hole Binary Candidates}}.
\newblock \emph{\bibinfo{journal}{Astrophys. J.}}
  \textbf{\bibinfo{volume}{856}}, \bibinfo{pages}{42} (\bibinfo{year}{2018}).

\bibitem{Sesana:2009:L129}
\bibinfo{author}{Sesana, A.}, \bibinfo{author}{Gair, J.},
  \bibinfo{author}{Mandel, I.} \& \bibinfo{author}{Vecchio, A.}
\newblock \bibinfo{title}{{Observing Gravitational Waves from the First
  Generation of Black Holes}}.
\newblock \emph{\bibinfo{journal}{Astrophys. J. Lett.}}
  \textbf{\bibinfo{volume}{698}}, \bibinfo{pages}{L129--L132}
  (\bibinfo{year}{2009}).

\bibitem{Mingarelli:2013}
\bibinfo{author}{{Mingarelli}, C.~M.~F.}, \bibinfo{author}{{Sidery}, T.},
  \bibinfo{author}{{Mandel}, I.} \& \bibinfo{author}{{Vecchio}, A.}
\newblock \bibinfo{title}{{Characterizing gravitational wave stochastic
  background anisotropy with pulsar timing arrays}}.
\newblock \emph{\bibinfo{journal}{\prd}} \textbf{\bibinfo{volume}{88}},
  \bibinfo{pages}{062005} (\bibinfo{year}{2013}).
\newblock \eprint{1306.5394}.

\bibitem{Ravi:2014:56}
\bibinfo{author}{Ravi, V.}, \bibinfo{author}{Wyithe, J. S.~B.},
  \bibinfo{author}{Shannon, R.~M.}, \bibinfo{author}{Hobbs, G.} \&
  \bibinfo{author}{Manchester, R.~N.}
\newblock \bibinfo{title}{{Binary supermassive black hole environments diminish
  the gravitational wave signal in the pulsar timing band}}.
\newblock \emph{\bibinfo{journal}{Mon. Not. R. Astron. Soc.}}
  \textbf{\bibinfo{volume}{442}}, \bibinfo{pages}{56--68}
  (\bibinfo{year}{2014}).

\bibitem{Roebber:2016:163}
\bibinfo{author}{Roebber, E.}, \bibinfo{author}{Holder, G.},
  \bibinfo{author}{Holz, D.~E.} \& \bibinfo{author}{Warren, M.}
\newblock \bibinfo{title}{{Cosmic Variance in the Nanohertz Gravitational Wave
  Background}}.
\newblock \emph{\bibinfo{journal}{Astrophys. J.}}
  \textbf{\bibinfo{volume}{819}}, \bibinfo{pages}{163} (\bibinfo{year}{2016}).

\bibitem{Habouzit:2016:1901}
\bibinfo{author}{Habouzit, M.} \emph{et~al.}
\newblock \bibinfo{title}{{Black hole formation and growth with non-Gaussian
  primordial density perturbations}}.
\newblock \emph{\bibinfo{journal}{Mon. Not. R. Astron. Soc.}}
  \textbf{\bibinfo{volume}{456}}, \bibinfo{pages}{1901--1912}
  (\bibinfo{year}{2016}).

\bibitem{Rosas-Guevara:2016:190}
\bibinfo{author}{{Rosas-Guevara}, Y.} \emph{et~al.}
\newblock \bibinfo{title}{{Supermassive black holes in the EAGLE Universe.
  Revealing the observables of their growth}}.
\newblock \emph{\bibinfo{journal}{\mnras}} \textbf{\bibinfo{volume}{462}},
  \bibinfo{pages}{190--205} (\bibinfo{year}{2016}).
\newblock \eprint{1604.00020}.

\bibitem{Kelley:2017:3131}
\bibinfo{author}{{Kelley}, L.~Z.}, \bibinfo{author}{{Blecha}, L.} \&
  \bibinfo{author}{{Hernquist}, L.}
\newblock \bibinfo{title}{{Massive black hole binary mergers in dynamical
  galactic environments}}.
\newblock \emph{\bibinfo{journal}{\mnras}} \textbf{\bibinfo{volume}{464}},
  \bibinfo{pages}{3131--3157} (\bibinfo{year}{2017}).
\newblock \eprint{1606.01900}.

\bibitem{Kelley:2017:4508}
\bibinfo{author}{Kelley, L.~Z.}, \bibinfo{author}{Blecha, L.},
  \bibinfo{author}{Hernquist, L.}, \bibinfo{author}{Sesana, A.} \&
  \bibinfo{author}{Taylor, S.~R.}
\newblock \bibinfo{title}{{The gravitational wave background from massive black
  hole binaries in Illustris: spectral features and time to detection with
  pulsar timing arrays}}.
\newblock \emph{\bibinfo{journal}{Mon. Not. R. Astron. Soc.}}
  \textbf{\bibinfo{volume}{471}}, \bibinfo{pages}{4508--4526}
  (\bibinfo{year}{2017}).

\bibitem{Kelley:2018:964}
\bibinfo{author}{Kelley, L.~Z.}, \bibinfo{author}{Blecha, L.},
  \bibinfo{author}{Hernquist, L.}, \bibinfo{author}{Sesana, A.} \&
  \bibinfo{author}{Taylor, S.~R.}
\newblock \bibinfo{title}{{Single sources in the low-frequency gravitational
  wave sky: properties and time to detection by pulsar timing arrays}}.
\newblock \emph{\bibinfo{journal}{Mon. Not. R. Astron. Soc.}}
  \textbf{\bibinfo{volume}{477}}, \bibinfo{pages}{964--976}
  (\bibinfo{year}{2018}).

\bibitem{Tremmel:2017:1121}
\bibinfo{author}{Tremmel, M.} \emph{et~al.}
\newblock \bibinfo{title}{{The Romulus cosmological simulations: a physical
  approach to the formation, dynamics and accretion models of SMBHs}}.
\newblock \emph{\bibinfo{journal}{Mon. Not. R. Astron. Soc.}}
  \textbf{\bibinfo{volume}{470}}, \bibinfo{pages}{1121--1139}
  (\bibinfo{year}{2017}).

\bibitem{Detweiler:1983:67}
\bibinfo{author}{Detweiler, S.} \& \bibinfo{author}{Ove, R.}
\newblock \bibinfo{title}{{Instability of some black holes}}.
\newblock \emph{\bibinfo{journal}{Physical Review Letters (ISSN 0031-9007)}}
  \textbf{\bibinfo{volume}{51}}, \bibinfo{pages}{67--70}
  (\bibinfo{year}{1983}).

\bibitem{Gualandris:2005:845}
\bibinfo{author}{Gualandris, A.}, \bibinfo{author}{Colpi, M.},
  \bibinfo{author}{Portegies~Zwart, S.} \& \bibinfo{author}{Possenti, A.}
\newblock \bibinfo{title}{{Has the Black Hole in XTE J1118+480 Experienced an
  Asymmetric Natal Kick?}}
\newblock \emph{\bibinfo{journal}{Astrophys. J.}}
  \textbf{\bibinfo{volume}{618}}, \bibinfo{pages}{845--851}
  (\bibinfo{year}{2005}).

\bibitem{Cicone:2018:143}
\bibinfo{author}{{Cicone}, C.} \emph{et~al.}
\newblock \bibinfo{title}{{ALMA [C I]$^{3}$ P $_{1}$-$^{3}$ P $_{0}$
  Observations of NGC 6240: A Puzzling Molecular Outflow, and the Role of
  Outflows in the Global {$\alpha$} $_{CO}$ Factor of (U)LIRGs}}.
\newblock \emph{\bibinfo{journal}{\apj}} \textbf{\bibinfo{volume}{863}},
  \bibinfo{pages}{143} (\bibinfo{year}{2018}).
\newblock \eprint{1807.06015}.

\bibitem{Rodriguez:2006:49}
\bibinfo{author}{{Rodriguez}, C.} \emph{et~al.}
\newblock \bibinfo{title}{{A Compact Supermassive Binary Black Hole System}}.
\newblock \emph{\bibinfo{journal}{\apj}} \textbf{\bibinfo{volume}{646}},
  \bibinfo{pages}{49--60} (\bibinfo{year}{2006}).
\newblock \eprint{astro-ph/0604042}.

\bibitem{Foord:2019}
\bibinfo{author}{Foord, A.}
\newblock \bibinfo{title}{{\textit{et al.} A Bayesian Analysis of SDSS
  J0914+0853, a Low-Mass Dual AGN Candidate (\textit{subm. to AAS Journals})}}.

\bibitem{Koss:2012:4264}
\bibinfo{author}{Koss, M.}
\newblock \bibinfo{title}{{a Census of Binary AGN, Feedback, and Obscured Black
  Hole Growth in the Local Universe}}.
\newblock \emph{\bibinfo{journal}{Chandra proposal ID {\#}14700336}}
  \textbf{\bibinfo{volume}{-1}}, \bibinfo{pages}{4264} (\bibinfo{year}{2012}).

\bibitem{U13}
\bibinfo{author}{{U}, V.} \emph{et~al.}
\newblock \bibinfo{title}{{The Inner Kiloparsec of Mrk 273 with Keck Adaptive
  Optics}}.
\newblock \emph{\bibinfo{journal}{\apj}} \textbf{\bibinfo{volume}{775}},
  \bibinfo{pages}{115} (\bibinfo{year}{2013}).
\newblock \eprint{1307.8440}.

\bibitem{U19}
\bibinfo{author}{{U}, V.} \emph{et~al.}
\newblock \bibinfo{title}{{Keck OSIRIS AO LIRG Analysis (KOALA): Feedback in
  the Nuclei of Luminous Infrared Galaxies}}.
\newblock \emph{\bibinfo{journal}{\apj}} \textbf{\bibinfo{volume}{871}},
  \bibinfo{pages}{166} (\bibinfo{year}{2019}).

\bibitem{Donley:2012:142}
\bibinfo{author}{Donley, J.~L.} \emph{et~al.}
\newblock \bibinfo{title}{{Identifying Luminous Active Galactic Nuclei in Deep
  Surveys: Revised IRAC Selection Criteria}}.
\newblock \emph{\bibinfo{journal}{Astrophys. J.}}
  \textbf{\bibinfo{volume}{748}}, \bibinfo{pages}{142} (\bibinfo{year}{2012}).

\bibitem{Kirkpatrick:2017:111}
\bibinfo{author}{{Kirkpatrick}, A.} \emph{et~al.}
\newblock \bibinfo{title}{{The AGN-Star Formation Connection: Future Prospects
  with JWST}}.
\newblock \emph{\bibinfo{journal}{\apj}} \textbf{\bibinfo{volume}{849}},
  \bibinfo{pages}{111} (\bibinfo{year}{2017}).
\newblock \eprint{1706.09056}.

\bibitem{Scoville:2015:70}
\bibinfo{author}{{Scoville}, N.} \emph{et~al.}
\newblock \bibinfo{title}{{ALMA Imaging of HCN, CS, and Dust in Arp 220 and NGC
  6240}}.
\newblock \emph{\bibinfo{journal}{\apj}} \textbf{\bibinfo{volume}{800}},
  \bibinfo{pages}{70} (\bibinfo{year}{2015}).
\newblock \eprint{1412.5183}.

\bibitem{Saito:2018:L52}
\bibinfo{author}{{Saito}, T.} \emph{et~al.}
\newblock \bibinfo{title}{{Imaging the molecular outflows of the prototypical
  ULIRG NGC 6240 with ALMA}}.
\newblock \emph{\bibinfo{journal}{\mnras}} \textbf{\bibinfo{volume}{475}},
  \bibinfo{pages}{L52--L56} (\bibinfo{year}{2018}).
\newblock \eprint{1712.07660}.

\bibitem{Treister:2018:83}
\bibinfo{author}{Treister, E.} \emph{et~al.}
\newblock \bibinfo{title}{{Optical, Near-IR, and Sub-mm IFU Observations of the
  Nearby Dual Active Galactic Nuclei MRK 463}}.
\newblock \emph{\bibinfo{journal}{Astrophys. J.}}
  \textbf{\bibinfo{volume}{854}}, \bibinfo{pages}{83} (\bibinfo{year}{2018}).

\bibitem{Scoville:2017:66}
\bibinfo{author}{{Scoville}, N.} \emph{et~al.}
\newblock \bibinfo{title}{{ALMA Resolves the Nuclear Disks of Arp 220}}.
\newblock \emph{\bibinfo{journal}{\apj}} \textbf{\bibinfo{volume}{836}},
  \bibinfo{pages}{66} (\bibinfo{year}{2017}).
\newblock \eprint{1605.09381}.

\bibitem{Sakamoto:2017:14}
\bibinfo{author}{{Sakamoto}, K.} \emph{et~al.}
\newblock \bibinfo{title}{{Resolved Structure of the Arp 220 Nuclei at
  {$\lambda$} $\approx$ 3 mm}}.
\newblock \emph{\bibinfo{journal}{\apj}} \textbf{\bibinfo{volume}{849}},
  \bibinfo{pages}{14} (\bibinfo{year}{2017}).
\newblock \eprint{1709.08537}.

\bibitem{Barcos-Munoz:2018:L28}
\bibinfo{author}{{Barcos-Mu{\~n}oz}, L.} \emph{et~al.}
\newblock \bibinfo{title}{{Fast, Collimated Outflow in the Western Nucleus of
  Arp 220}}.
\newblock \emph{\bibinfo{journal}{\apjl}} \textbf{\bibinfo{volume}{853}},
  \bibinfo{pages}{L28} (\bibinfo{year}{2018}).
\newblock \eprint{1712.06381}.

\bibitem{Sakamoto:2014:90}
\bibinfo{author}{{Sakamoto}, K.}, \bibinfo{author}{{Aalto}, S.},
  \bibinfo{author}{{Combes}, F.}, \bibinfo{author}{{Evans}, A.} \&
  \bibinfo{author}{{Peck}, A.}
\newblock \bibinfo{title}{{An Infrared-luminous Merger with Two Bipolar
  Molecular Outflows: ALMA and SMA Observations of NGC 3256}}.
\newblock \emph{\bibinfo{journal}{\apj}} \textbf{\bibinfo{volume}{797}},
  \bibinfo{pages}{90} (\bibinfo{year}{2014}).
\newblock \eprint{1403.7117}.

\bibitem{Treister:2019}
\bibinfo{author}{Treister, E.}, \bibinfo{author}{Messias, H.},
  \bibinfo{author}{Privon, C.} \& \bibinfo{author}{Koss, M.}
\newblock \bibinfo{title}{{The Molecular Gas in the NGC6240 Merging Galaxy
  System at the Highest Spatial Resolution (\textit{submitted})}}.

\bibitem{Aalto:1995:369}
\bibinfo{author}{{Aalto}, S.}, \bibinfo{author}{{Booth}, R.~S.},
  \bibinfo{author}{{Black}, J.~H.} \& \bibinfo{author}{{Johansson}, L.~E.~B.}
\newblock \bibinfo{title}{{Molecular gas in starburst galaxies: line
  intensities and physical conditions}}.
\newblock \emph{\bibinfo{journal}{\aap}} \textbf{\bibinfo{volume}{300}},
  \bibinfo{pages}{369} (\bibinfo{year}{1995}).

\bibitem{Downes:1998:615}
\bibinfo{author}{{Downes}, D.} \& \bibinfo{author}{{Solomon}, P.~M.}
\newblock \bibinfo{title}{{Rotating Nuclear Rings and Extreme Starbursts in
  Ultraluminous Galaxies}}.
\newblock \emph{\bibinfo{journal}{\apj}} \textbf{\bibinfo{volume}{507}},
  \bibinfo{pages}{615--654} (\bibinfo{year}{1998}).
\newblock \eprint{astro-ph/9806377}.

\bibitem{Papadpoulos:2012:2601}
\bibinfo{author}{{Papadopoulos}, P.~P.} \emph{et~al.}
\newblock \bibinfo{title}{{The molecular gas in luminous infrared galaxies - I.
  CO lines, extreme physical conditions and their drivers}}.
\newblock \emph{\bibinfo{journal}{\mnras}} \textbf{\bibinfo{volume}{426}},
  \bibinfo{pages}{2601--2629} (\bibinfo{year}{2012}).
\newblock \eprint{1109.4176}.

\end{thebibliography}

\end{document}